# Polarization rotation in $Bi_4Ti_3O_{12}$ by isovalent doping at the fluorite sublattice

Kevin Co,[1] Fu-Chang Sun,[1] S. Pamir Alpay,[1,2] and Sanjeev K. Nayak[1,a]

[1] *Department of Materials Science and Engineering and Institute of Materials Science*
*University of Connecticut, Storrs, CT 06269 – USA*

[2] *Department of Physics, University of Connecticut, Storrs, CT 06269 – USA*

## ABSTRACT

Bismuth titanate, $Bi_4Ti_3O_{12}$ (BiT), is a complex layered ferroelectric material that is composed of three perovskite-like units and one fluorite-like unit stacked alternatively along the *c*-direction. The ground state crystal structure is monoclinic with the spontaneous polarization (~50 $\mu C/cm^2$) along the in-plane *b*-direction. BiT typically grows along the *c*-direction in thin film form and having the polarization vector aligned with the growth orientation can be beneficial for several potential device applications. It is well known that judicious doping of ferroelectrics is an effective method in adjusting the magnitude and the orientation of the spontaneous polarization. Here, we show using first-principles density functional theory and a detailed phonon analysis that Bi atoms in the fluorite-like layers have significantly more impact on the magnitude and orientation of the spontaneous polarization vector as compared to the perovskite-like layer. The low energy hard phonon modes are characterized by fluorite-like layers experiencing transverse displacements and large changes in Born effective charges on Bi atoms. Thus, the breaking of symmetry caused by doping of Bi sites within the fluorite-like layer leads to the formation of uncancelled permanent dipole moments along the *c*-direction. This provides an opportunity for doping the Bi site in the fluorite-like layer. Isovalent dopants P, As, and Sb were studied. P is found to be most effective in the reorientation of the spontaneous polarization. It leads to a three-fold enhancement of the *c*-component of polarization and to a commensurate rotation of the spontaneous polarization vector by 36.2° towards the *c*-direction.

---

[a] Corresponding author: sanjeev.nayak@uconn.edu

# I. Introduction

Bismuth titanate, $Bi_4Ti_3O_{12}$ (BiT), is a ferroelectric material, notable for its layered monoclinic ground state crystal structure of the space group $B1a1$ ($P_c$ according to the international space group classification). The crystal structure belongs to the Aurivillius family and contains stacks of perovskite-like (PL) and fluorite-like (FL) units with a chemical formula $(Bi_2O_2)(Bi_{n-1}Ti_nO_{3n+1})$, $n \geq 1$ [1,2]. The crystal structure of BiT has alternating FL $(Bi_2O_2)^{2+}$ and PL $(Bi_{n-1}Ti_nO_{3n+1})^{2-}$ sub-layers stacked along the $c$-axis with the number of perovskite-like layers, $n = 3$. The Curie temperature of BiT is measured as $T_C = 675$ °C [2]. The ferroelectric $B1a1$ structure transforms to a paraelectric tetragonal crystal structure above $T_C$ with space group $I4/mmm$. Figure 1 shows the schematic representation of the BiT crystal structure and the FL and PL layers in the paraelectric and ferroelectric states. The spontaneous polarization of BiT at 25 °C is measured to be $\sim 50$ μC/cm$^2$ which lies mostly parallel to the $b$-axis and a small component of 4 μC/cm$^2$ appears along the $c$-direction [3,4]. In comparison, the polarization value of prototypical perovskite ferroelectric compounds $BaTiO_3$ and $PbTiO_3$ are 28.4 μC/cm$^2$ and 66 μC/cm$^2$, respectively [5], [6]. Due to relatively large $T_C$ and its lead-free chemistry, BiT is an attractive ferroelectric ceramic compound that may find applications in many functional devices, including dynamic and non-volatile memories [6]–[7], electrically tunable capacitors for telecommunications [9], pyroelectric sensors, piezoelectric actuators, and in high-temperature waste heat interconversion [9]–[15]. Layered complex oxides such as BiT are also promising materials for picoscale engineering because of a well-defined compositional and structural interface which can be controlled experimentally in a very precise manner [10] .

Epitaxial BiT samples deposited by pulsed laser deposition on a variety of substrates (such as MgO, $LaAlO_3$, $SrTiO_3$, $MgAl_2O_4$, and Si) typically grow preferentially along the crystallographic $c$-axis [4,11]. This is due to the low surface energy of (001) planes leading to the nucleation of (001) oriented platelets by an Ostwald ripening process which grow into highly textured grains to minimize grain boundary energy [5]. Biaxial internal strains resulting from lattice mismatch in epitaxial films result in variations in the Curie transition temperature, phase transformation mode, and phase stabilities [12–19]. This misfit strain can be adjusted by varying the substrate material and by film thickness. While epitaxial strain can result in polarization rotation, such a phenomenon has not been observed in BiT thin films [20]. Previous studies on epitaxially grown stoichiometric thin films indicate that the principal component of polarization remains parallel to the growth plane, with the major component in the $b$-direction [4,11].

Chemical alterations through judicious doping produce significant variations in the properties of ferroelectrics. Hitherto, chemical manipulation is conceived for perovskite cation sites only. Doping of the Ti site in BiT in the PL layer with La and V has led to increased chemical stability of oxygen vacancies, resulting in reduced ionic and hole conductivity along the $ab$ axis and highly desirable polarization-fatigue resistance [21,22], [23], [24,25]. Niobium doping at the Ti site is found to improve polarization fatigue resistance which also led to an increase in the remnant polarization through tilting of the oxygen octahedra [7], [26]. Similarly, V and W additions to the Ti site in the PL layer produces an enhancement in the remnant polarization which is believed to be related to domain phenomena [27]. Neodymium addition has been shown to result in lower remnant polarization and leakage current and tantalum doping results in lower dielectric loss [28], [29]. Isovalent doping has also been explored in experiments, but is limited to PL sites [30,31]. We note that developing ferroelectric functionalities by doping with transition metals has been explored by first principles methods but, consistent with experimental work to date, only the Ti site in the PL layer was investigated [32,33].

To date, experimental and computational studies on BiT and other ferroelectric systems clearly indicate that variations in the polarization response can be achieved by introducing chemical and structural perturbations. That being said, the possibility of reorientation of the polarization vector along the longer crystallographic $c$-direction in BiT (and other Aurivillius phases) has not been explored at all. Such a study could be particularly valuable in the context of non-volatile memory devices since it is preferable for polarization switching to develop normal to the probe direction. This would lead to a higher resolution of detecting the switching states during read and write operations [7]. It has been possible to tune the polarization rotation in artificially grown superlattice thin films, emphasizing the tunability of properties by precise control of the interface structure [34,35]. Bi site doping within PL layers with La and Nb have been carried out in a combined theoretical and experimental study and there were no significant changes in the $c$-component of polarization [36].

Dion-Jacobson and Ruddelson-Popper phases are layered perovskite structures, similar to Aurivillius type materials, characterized by perovskite units and interwoven interfacial cation layers along [001] and existing in a tetragonal phase [37–39]. In these structures, ferroelectricity is improper in its character, originating from rotations of oxygen octahedra. In contrast, the $Bi_2O_2$ layer of the Aurivillius phase stabilizes a monoclinic structure upon cooling and ferroelectricity in BiT is of mixed proper-improper character [40]. Combined with the fact that the phonon density of states at low energies have maximum contributions from the Bi atoms [41], it is evident that chemical doping at the Bi sites in the $Bi_2O_2$ layer holds much promise in altering ferroelectric properties of BiT and BiT-like ferroelectrics. Here, we focus on the chemical manipulation of these Bi sites in the FL layer to achieve out-of-plane orientation of spontaneous polarization. Our results obtained from first-principles theory and phonon analysis indicate that Bi atoms in the fluorite-like layer have the largest contributions to the dipole moment for low energy hard phonon modes. This provides a unique opportunity to adjust the magnitude and the orientation of the spontaneous polarization vector by replacing the Bi ions in the FL layers with isovalent atoms such as P, As, and S. We show here that out of these P is found to be most promising in rotating the polarization vector from *b*-axis towards the *c*-direction and in modulating the band gap, which is attractive for opto-electronic applications. These results have significant implications in designing devices with BiT and related materials.

## II. Computational Approach

First principles calculations are performed with density functional theory using the plane wave pseudopotential method [42], [43]. The relaxed lattice parameters for the ferroelectric monoclinic $B1a1$ and paraelectric tetragonal $I4/mmm$ are taken from Ref. [41]. The lattice parameters for BiT in the monoclinic phase are $a = 5.49$ Å, $b = 5.53$ Å, $c = 16.88$ Å, and $\alpha = 80.61°$, $\beta = \gamma = 90°$, and for tetragonal phase, $a = b = 3.85$ Å, $c = 33.20$ Å, and $\alpha = \beta = \gamma = 90°$. The lattice vectors and the atomic positions of the doped models are thoroughly optimized such that there are no residual interatomic forces. Non-spin polarized calculations are performed with generalized gradient approximation with Perdew-Burke-Ernzerhof (PBE) parametrization [44] as the exchange-correlation functional. The core and valence electrons are treated with the projector augmented wave method [45]. The kinetic energy cutoff for the plane waves is set to 500 eV. The integrations in the Brillouin zone is performed in a discretized Monkhorst-Pack [46] $k$-points mesh of $5 \times 5 \times 3$ for the geometrical optimization and $8 \times 8 \times 8$ for the self-consistent field. Geometrical optimization for the models is carried out with the tolerance for total energy convergence set to $10^{-6}$ eV. Born effective charge (BEC) is calculated by the linear response method using the density-functional perturbation theory (DFPT) [47]. The calculations are performed with the Vienna *ab initio* simulation package (VASP) [48–51]. The phonon modes are taken from a previous study [41] where Phonopy [52] was used in the computations.

## III. Spontaneous Polarization

According to the modern theory of polarization, the electric polarization is a differential property which is defined between the ferroelectric and the paraelectric phases of the same material [53–55]. The paraelectric reference is derived from the low-temperature phase through a continuous transformation variable which preserves the insulating character throughout the transformation. In prototype ferroelectric perovskites such as BaTiO$_3$, the transformation path is through a single symmetry breaking transverse-optic phonon mode called the soft mode which is a characteristic of a structural phase transition [56]. BiT exhibits a more complicated phase transformation path compared to BaTiO$_3$ where multiple soft-phonon modes result in several intermediary phases through which the phase transition occurs. This is illustrated in Figure 2 with data taken from Ref. [51]. Therefore, the paraelectric reference state must be represented by a commensurate centrosymmetric crystal structure following the principles of the modern theory of polarization. The lack of a precise transformation path for the change in crystal structure available from experiments produces only a linear mapping between the structures for DFT calculations, which is often inelegant. Particular to the layered systems, there is an additional difficulty stemming from the fact that the length on one of the lattice vectors is much larger than others. This leads to smaller values of polarization quantum in the transverse directions, but the polarization components are much larger than the polarization quantum, making it difficult to track the polarization change from linear atomic displacements. Given the fact that the lattice vectors and crystal volume vary between the reference states, the polarization quantum does not remain constant throughout the intermediate images, which generates further problems in modeling.

As such, we have computed the polarization from the product of Born effective charges (BECs) and the ionic displacement vectors, taken by matching the shortest displacements between the ions of reference structures for our studies on doped BiT. The polarization component obtained from this approach is slightly larger than the reported experimental values. Since the initial and the final states are critical for the estimation of spontaneous polarization, the analysis is still useful with respect to the ferroelectric polarization, but the transformation path may not be explicitly derived from such an approach.

BECs and the atomic displacement vectors are central quantities that assist in the computation of the atomic dipole moment. Very generally, the spontaneous polarization is the integrated dipole moment over a unit volume, i.e., the dipole moment density. BEC is a tensorial quantity which is defined as the change in the polarization along a direction $i$ induced by an atomic displacement along a direction $j$ in the absence of external applied electric field. It is given by [53,59]:

$$Z^*_{ij} = \frac{\Omega}{e} \frac{\delta P_i}{\delta d_j},\qquad(1)$$

where, $\Omega$ and $e$ are the unit cell volume and the electronic charge, respectively. The BECs computed for different atoms in the $B1a1$ structure of BiT are shown in Table I for PL and FL layers. The different atom types are indexed from 1 to 6. The symmetrically equivalent atoms are primed ($'$). The nomenclature is kept consistent with Ref. [60] to allow for a direct comparison. The BECs calculated here match with the ones reported in [60] with minor deviations that can be attributed to differences in the lattice constants. To compute the displacement vectors between ferroelectric and paraelectric BiT, it is necessary to orient the supercells such that the displacements can be matched one-to-one (see Figure 1). This is achieved by applying the transformation matrices $T_m$ and $T_t$ to the unit cells of the monoclinic and the tetragonal lattices, respectively:

$$T_m = \begin{bmatrix} 1 & 0 & 0 \\ 0 & 1 & 0 \\ 0 & 0 & 2 \end{bmatrix} \text{ and } T_t = \begin{bmatrix} \sqrt{2}\cos\left(\frac{\pi}{4}\right) & -\sqrt{2}\sin\left(\frac{\pi}{4}\right) & 0 \\ \sqrt{2}\sin\left(\frac{\pi}{4}\right) & \sqrt{2}\cos\left(\frac{\pi}{4}\right) & 0 \\ 0 & 0 & 1 \end{bmatrix} \qquad (2)$$

Atoms between the two structures are then matched first by their positions along the $c$ axis ([001]) and the displacement vectors along the $ab$ plane are calculated as the shortest displacement between the periodic structures. With the mapping as defined above, the $b$- and $c$- components of spontaneous polarization are calculated as 52.25 $\mu C/cm^2$ and 7.88 $\mu C/cm^2$, respectively, which is in good agreement with previous DFT calculations and experimental literature [60], [4]. The macroscopic value of the spontaneous polarization for the paraelectric $I4/mmm$ structure is zero.

For the $B1a1$ structure, the off-center displacements of charges (ODCs) are central to the appearance of permanent electric dipoles. In Table II, the average ODC and average cation BEC are shown for different cation positions in the $B1a1$ structure. One clearly observes that the $b$- and $c$-components of the ODC is the largest for the $Bi_{FL}$ atoms. The numerical averages indicate that Bi ions have non-negligible contribution to polarization owing to large ODC and BEC as compared to Ti atoms. It is also worth pointing out that the number of neighboring O atoms for Bi in FL and Bi in PL layers are 8 and 6, respectively, for the $B1a1$ structure, while these are 12 and 8 for the $I4/mmm$ structure, representing a four and two atom change in nearest neighbors in FL and PL. Further support for the doping of the cation site in the FL layer comes from a detailed phonon analysis which is described in the next section.

## IV. Phonon Analysis

Phonon analysis has been systematically used to study phase transformation and associated polarization variations in ferroelectrics [61]. Phonon modes, as used in this study, are a better approximation for accounting the ionic displacement vectors than a linear mapping between ferro- and paraelectric structures. Via this methodology, mapping the reference structures is not optimal due to the harmonic approximation; however, it provides a view of atomic contribution to dipole moments within the ferroelectric structure.

Previous studies on phonon analysis in BiT have been focused on soft phonons and structural transformation pathways for the tetragonal high-symmetry phase [58,62]. However, such approaches have not been utilized in the description of the spontaneous polarization vector. In a recent experimental study, using THz spectroscopy, three phonon modes in BiT were resolved having energy 0.68 THz, 0.86 THz, and 0.96 THz [41]. Prior to this report, experiments reported only one phonon mode at 0.83 THz [24,63]. DFT combined with a frozen phonon method showed a normal band structure (no negative energies were observed) for the $B1a1$ structure, emphasizing that this structure is dynamically stable. To gain insight into the phase transformation characteristics, the ground state monoclinic lattice is subjected here to a transverse strain that results in change of the monoclinic angle. DFPT calculations for the ground state $B1a1$ structure provided the energies of phonons at the $\Gamma$-point which were in the following sequence: $\omega_1 = 1.108$ THz > $\omega_2 = 1.267$ THz > $\omega_3 = 1.299$ THz > $\omega_4 = 1.604$ THz.

The energies of the four phonon modes as a function of their amplitude are positive and proportional to amplitude. At 0.04% transverse strain condition, two modes, $\omega_2$ and $\omega_3$ lower the total energy with an increase in the amplitude, see Figure 3. The minimum energy obtained is 27.6 meV and 25.1 meV below the reference ground state energy for $\omega_2$ and $\omega_3$, respectively. This behavior is a characteristic of soft modes that trigger a structural phase transition [56]. The two soft modes involve transverse displacement of FL blocks out-of-phase with the PL block (consistent with the rigid-layer mode defined by Machado *et al.* in Ref. [62]) for the tetragonal structures and also have components of rotation within the perovskite octahedra. The soft modes could be interpreted as the restoring path to the ground state when the $B1a1$ structure is perturbed by transverse strain along the $bc$-plane. The atomic displacements for the modes $\omega_2$, $\omega_3$ and $\omega_4$ are shown by arrows in Figure 4. The $\omega_1$ hard mode is characterized by transverse displacement of FL and PL layers along the non-polar $a$ axis and negligible displacement along the polar axes, whereas the hard mode, $\omega_4$, exhibits transverse displacements between layers along the $c$ axis.

Difference in the BEC of atoms ($dZ_i^*$) obtained for the four phonon-displaced structure and the $B1a1$ reference structure as well as the atomic displacement of the phonon modes ($d\eta_i$) are then used to compute the layer resolved dipole moment (LRDM) given by the expression:

$$\text{LRDM} = \sum d\eta_i dZ_i^* . \qquad (3)$$

The $a$-, $b$- and $c$-components of LRDM are listed in Table III. The LRDM provides an insightful view of uncancelled ionic dipoles. It is observed that the FL layer has a significant influence on LRDM for the hard modes $\omega_1$ and $\omega_4$, implying greater contribution from Bi atoms, which surpass the impact of Ti atoms. This observation further reinforces the idea of targeting the FL layer Bi atoms for controlling the properties in BiT and, more generally, the transition metal in the FL in Aurivillius type ceramics.

The changes in the atomic dipole contributions per structural unit (FL and PL units) are summarized in Figure 5. Here, it is clear that the involvement of the PL layer is larger for the soft modes $\omega_2$ and $\omega_3$. This is expected through the phenomenological understanding of polarization variations occurring via soft mode-driven displacements [56]. For the hard mode $\omega_4$, which is characterized by transverse displacement of the FL and PL layers along the $c$-direction, the LRDM of the FL layer is greater than that of the PL and is mostly due to the presence of Bi atoms. These results clearly indicate that doping the Bi site in FL layers should produce significant alterations in the magnitude of the dipole moments. Considering that there are transverse displacements associated with the hard mode $\omega_4$ (see Figure 4c and 5), one would also expect a reorientation of the polarization vector towards the *c*-direction.

## V. Isovalent Doping

It is emphasized thus far that the Bi atoms in the FL layer effectively carry the control of polarization response of BiT through transverse displacements along the *c*-direction. An effective way of inducing specific changes in crystal structure, phonon modes and the polarization is by doping. We studied the polarization changes due to doping for the Bi atoms in the FL layer by isovalent elements to Bi. Group 15 elements P, As, and Sb in the periodic table are chosen as dopants, assuming these preserve the host crystal structure. The doping concentration corresponds to 6.25%, which can be practically achievable by non-equilibrium synthesis methods such as pulsed laser deposition. Nitrogen is excluded in the study due to the large size difference between N and Bi atoms. Coupled with the fact that N has a relatively high electronegativity, adding N cannot be expected to preserve the local symmetry, hence the FL structure, and the FL-PL layer arrangement necessary for the formation of a Aurivillius phase.

In these studies, the supercell geometry is optimized with respect to both atomic positions and the supercell volume to minimize residual internal forces. The polarization components, absolute value of the polarization, the angle of polarization with respect to the *c*-axis, and the band gap for different $a_i$ conditions are given in Table IV. The partial electronic density of states (PDOS) and the total electronic density of states (TDOS) are shown for P, As, and Sb doped systems in Figures 6(a)-(c) and (d)-(f), respectively. The dopant atoms hybridize across the energy range of valence band of BiT. Unlike Sb, the low-lying states of P and As near the conduction band results in an effective change in the band gap, which changes systematically as one goes down the group for the choice of dopants.The wide band gap characteristics are preserved in the doped BiT. P and As doping results in a 36.2% and 17.5% reduction in the band gap, respectively, while Sb doping results in a 2.0% increase as obtained from GGA calculations. It must be noted that GGA underestimates the band gap in semiconductors, however the trends across the band gap with doping is expected to hold true to the real system. The polarization change appears in the order of few percent as compared to pure BiT, with P resulting in a slightly larger value of polarization than undoped BiT. The polarization vector is observed to have a relatively larger *c*-component than that of the pure BiT. The polarization is rotated towards the *c*-direction for all three dopants, but the highest rotation is observed for the P doped cases, which corresponds to a three-fold enhancement of the *c*-component of polarization, and a commensurate rotation of the spontaneous polarization vector to 36.2° out-of-plane. As and Sb doping result in spontaneous polarization vectors of 27.2° and 27.9°, respectively and lower magnitudes of polarization, as compared to P doping.

# VI. Summary

Orientation control of the spontaneous polarization vector may provide greater functionality in ferroelectric films. Stabilization of orientational variants and new rotational phases in several ferroelectric systems including $BaTiO_3$ and $PbTiO_3$ have been investigated theoretically and experimentally for over two decades [12,64]. The same concepts can be utilized in BiT and other Aurivillius systems. In particular, BiT is a promising ferroelectric due to a large polarization and high Curie temperature but is limited in applications in thin film form due to its in-plane spontaneous polarization. In this report, we have shown using first principles methods and phonon studies that it is the Bi ion in the FL layer in the BiT crystal structure that contributes to transverse displacements for low energy hard phonon modes and large contributions to the phonon density of states. As such, we explored isovalent doping on the Bi FL site as a means of inducing uncancelled dipole moments transverse to the primary axis of polarization. Based on density functional theory, we have shown that P doping is a promising candidate for orientation control of the spontaneous polarization vector, resulting in a three-fold enhancement of the *c*-component of polarization and a commensurate rotation of the spontaneous polarization vector to 36.2° out-of-plane, while still maintaining the magnitude of polarization. As the $(Bi_2O_2)^{2+}$ layer is a key characteristic of all Aurivillius compounds, FL cation site doping can be an effective method of polarization orientation control in other such materials systems with similar crystal structures.

# Acknowledgements

The computational resources employed in this study were provided by High Performance Computing facilities at the University of Connecticut. SPA and SKN acknowledge the Extreme Science and Engineering Discovery Environment (XSEDE), which is supported by National Science Foundation grant number ACI-1548562. We express our appreciation for experimental feedback and discussions with D. Maurya (Virginia Tech) and S. Priya (Penn State). Discussions with K. C. Pitike, Oak Ridge National Laboratory, are also acknowledged.

# References

[1]  B. Aurivillius and P. H. Fang, Phys. Rev. **126**, 893 (1962).

[2]  A. Peláiz-Barranco and Y. González-Abreu, J. Adv. Dielectr. **3**, 1330003 (2013).

[3]  S. E. Cummins, J. Appl. Phys. **39**, 2268 (1968).

[4]  R. Ramesh, K. Luther, B. Wilkens, D. L. Hart, E. Wang, J. M. Tarascon, A. Inam, X. D. Wu, and T. Venkatesan, Appl. Phys. Lett. **57**, 1505 (1990).

[5]  H. H. Wieder, Phys. Rev. **99**, 1611 (1955).

[6]  V. G. Bhide, M. S. Hegde, and K. G. Deshmukh, J. Am. Ceram. Soc. **51**, 565 (1968).

[7]  O. Auciello, J. F. Scott, and R. Ramesh, Phys. Today **51**, 22 (1998).

[8]  J. F. Scott and C. A. Paz de Araujo, Science **246**, 1400 (1989).

[9]  G. Subramanyam, M. W. Cole, N. X. Sun, T. S. Kalkur, N. M. Sbrockey, G. S. Tompa, X. Guo, C. Chen, S. P. Alpay, G. A. Rossetti, K. Dayal, L. Q. Chen, and D. G. Schlom, J. Appl. Phys. **114**, 191301 (2013).

[10]  S. Ismail-Beigi, F. J. Walker, A. S. Disa, K. M. Rabe, and C. H. Ahn, Nat. Rev. Mater. **2**, 17060 (2017).

[11]  H. Buhay, S. Sinharoy, W. H. Kasner, M. H. Francombe, D. R. Lampe, and E. Stepke, Appl. Phys. Lett. **58**, 1470 (1991).

[12]  D. G. Schlom, L.-Q. Chen, C.-B. Eom, K. M. Rabe, S. K. Streiffer, and J.-M. Triscone, Annu. Rev. Mater. Res. **37**, 589 (2007).

[13]  N. A. Pertsev, A. G. Zembilgotov, and A. K. Tagantsev, Phys. Rev. Lett. **80**, 1988 (1998).

[14]  S. P. Alpay, J. Mantese, S. Trolier-Mckinstry, Q. Zhang, and R. W. Whatmore, Mater. Res. Bull. **39**, 1099 (2014).

[15]  I. B. Misirlioglu, S. P. Alpay, F. He, and B. O. Wells, J. Appl. Phys. **99**, 104103 (2006).

[16]  Q. Y. Qiu, R. Mahjoub, S. P. Alpay, and V. Nagarajan, Acta Mater. **58**, 823 (2010).

[17]  R. K. Vasudevan, H. Khassaf, Y. Cao, S. Zhang, A. Tselev, B. Carmichael, M. B. Okatan, S. Jesse, L. Q. Chen, S. P. Alpay, S. V Kalinin, and N. Bassiri-Gharb, Adv. Funct. Mater. **26**, 478 (2016).

[18]  A. Sharma, Z. G. Ban, S. P. Alpay, and J. V. Mantese, Appl. Phys. Lett. **85**, 985 (2004).

[19]  L. W. Martin and A. M. Rappe, Nat. Rev. Mater. **2**, 16087 (2016).

[20]  S. H. Shah and P. D. Bristowe, J. Phys. Condens. Matter **22**, 385902 (2010).

[21]  Y. Y. Yao, C. H. Song, P. Bao, D. Su, X. M. Lu, J. S. Zhu, and Y. N. Wang, J. Appl. Phys. **95**, 3126 (2004).

[22]  M. Takahashi, Y. Noguchi, and M. Miyayama, Jpn. J. Appl. Phys. **42**, 6222 (2003).

[23]  J. C. Bae, S. S. Kim, E. K. Choi, T. K. Song, W.-J. Kim, and Y.-I. Lee, Thin Solid Films **472**, 90 (2005).

[24]  B. H. Park, B. S. Kang, S. D. Bu, T. W. Noh, J. Lee, and W. Jo, Nature **401**, 682 (1999).

[25]  S. H. Shah and P. D. Bristowe, J. Phys. Condens. Matter **23**, 155902 (2011).

[26]  J. K. Kim, J. Kim, T. K. Song, and S. S. Kim, Thin Solid Films **419**, 225 (2002).

[27]  Y. Noguchi, I. Miwa, Y. Goshima, and M. Miyayama, Jpn. J. Appl. Phys. **39**, L1259 (2000).

[28]  T. Goto, Y. Noguchi, M. Soga, and M. Miyayama, Mater. Res. Bull. **40**, 1044 (2005).

[29]  S.-H. Hong, J. A. Horn, S. Trolier-McKinstry, and G. L. Messing, J. Mater. Sci. Lett. **19**, 1661 (2000).

[30]  W. Li, J. Gu, C. Song, D. Su, and J. Zhu, J. Appl. Phys. **98**, 114104 (2005).

[31]  O. Subohi, G. S. Kumar, M. M. Malik, and R. Kurchania, J. Phys. Chem. Solids **93**, 91 (2016).

[32]  A. Y. Birenbaum and C. Ederer, Phys. Rev. B **90**, 214109 (2014).

[33]  A. Y. Birenbaum, A. Scaramucci, and C. Ederer, Phys. Rev. B **95**, 104419 (2017).

[34]  J. Sinsheimer, S. J. Callori, B. Bein, Y. Benkara, J. Daley, J. Coraor, D. Su, P. W. Stephens, and M. Dawber, Phys. Rev. Lett. **109**, 167601 (2012).

[35]  H. Lu, X. Liu, J. D. Burton, C.-W. Bark, Y. Wang, Y. Zhang, D. J. Kim, A. Stamm, P. Lukashev, D. A. Felker, C. M. Folkman, P. Gao, M. S. Rzchowski, X. Q. Pan, C.-B. Eom, E. Y. Tsymbal, and A. Gruverman, Adv. Mater. **24**, 1209 (2012).

[36] A. Roy, R. Prasad, S. Auluck, and A. Garg, Appl. Phys. Lett. **102**, 182901 (2013).

[37] N. A. Benedek, Inorg. Chem. **53**, 3769 (2014).

[38] B. V. Beznosikov and K. S. Aleksandrov, Crystallogr. Rep. **45**, 792 (2000).

[39] S. N. Ruddlesden and P. Popper, Acta Crystallogr. **10**, 538 (1957).

[40] N. A. Benedek, J. M. Rondinelli, H. Djani, P. Ghosez, and P. Lightfoot, Dalton Trans. **44**, 10543 (2015).

[41] D. Maurya, A. Charkhesht, S. K. Nayak, F.-C. Sun, D. George, A. Pramanick, M.-G. Kang, H.-C. Song, M. M. Alexander, D. Lou, G. A. Khodaparast, S. P. Alpay, N. Q. Vinh, and S. Priya, Phys. Rev. B **96**, 134114 (2017).

[42] R. O. Jones and O. Gunnarsson, Rev. Mod. Phys. **61**, 689 (1989).

[43] C. W. Bark, Met. Mater. Int. **19**, 1361 (2013).

[44] J. Perdew, K. Burke, and M. Ernzerhof, Phys. Rev. Lett. **77**, 3865 (1996).

[45] P. E. Blöchl, Phys. Rev. B **50**, 17953 (1994).

[46] H. J. Monkhorst and J. D. Pack, Phys. Rev. B **13**, 5188 (1976).

[47] S. Baroni, S. de Gironcoli, A. Dal Corso, and P. Giannozzi, Rev. Mod. Phys. **73**, 515 (2001).

[48] G. Kresse and J. Furthmüller, Comp. Mater. Sci. **6**, 15 (1996).

[49] G. Kresse and J. Furthmüller, Phys. Rev. B **54**, 11169 (1996).

[50] G. Kresse and D. Joubert, Phys. Rev. B **59**, 1758 (1999).

[51] M. Gajdoš, K. Hummer, G. Kresse, J. Furthmüller, and F. Bechstedt, Phys. Rev. B **73**, 045112 (2006).

[52] A. Togo and I. Tanaka, Scr. Mater. **108**, 1 (2015).

[53] R. D. King-Smith and D. Vanderbilt, Phys. Rev. B **47**, 1651 (1993).

[54] R. Resta, Rev. Mod. Phys. **66**, 899 (1994).

[55] R. Resta and D. Vanderbilt, Top. Appl. Phys. **105**, 31 (2007).

[56] M. E. Lines and A. M. Glass, *Principles and Applications of Ferroelectrics and Related Materials* (Oxford University Press, Oxford, 2001).

[57] A. Shrinagar, A. Garg, R. Prasad, and S. Auluck, Acta Crystallogr. A **64**, 368 (2008).

[58] J. M. Perez-Mato, P. Blaha, K. Schwarz, M. Aroyo, D. Orobengoa, I. Etxebarria, and A. García, Phys. Rev. B **77**, 184104 (2008).

[59] N. A. Spaldin, J. Solid State Chem. **195**, 2 (2012).

[60] A. Roy, R. Prasad, S. Auluck, and A. Garg, J. Phys. Condens. Matter **22**, 165902 (2010).

[61] C.-Z. Wang, R. Yu, and H. Krakauer, Phys. Rev. B **54**, 11161 (1996).

[62] R. Machado, M. G. Stachiotti, R. L. Migoni, and A. Huanosta Tera, Phys. Rev. B **70**, 214112 (2004).

[63] D. Nuzhnyy, S. Kamba, P. Kužel, S. Veljko, V. Bovtun, M. Savinov, J. Petzelt, H. Amorín, M. E. V. Costa, A. L. Kholkin, Ph. Boullay, and M. Adamczyk, Phys. Rev. B **74**, 134105 (2006).

[64] D. G. Schlom, L. Chen, C. J. Fennie, V. Gopalan, D. A. Muller, X. Pan, R. Ramesh, and R. Uecker, Mater. Res. Bull. **39**, 118 (2014).

**Figure Captions**

**Figure 1**    (Color online) Schematic structure of the ground state $Ba_4Ti_3O_{12}$ projected along [010] direction (left) and the paraelectric *I4/mmm* structure (right) with zoomed projections of the local structural units (center). The ground state is a monoclinic cell with *B1a1* structure.

**Figure 2**    Group-subgroup graph connecting the *I4/mmm* and *B1a1* space groups with all intermediate subgroups (taken from Ref. [58]).

**Figure 3**    (Color online) Total energy of the supercell with respect to the ground state energy as a function of amplitude of atomic displacements along the modes $\omega_1 - \omega_4$ for ground state *B1a1* BiT (black lines) and that obtained by shearing the lattice to increase the monoclinic angle by 0.4% (red lines). The lowering of energy as a function of mode amplitude is characteristic property for soft modes. $\omega_2$ and $\omega_3$ qualify as soft phonons with energy 27.6 meV and 25.1 meV, respectively, below ground state at non-zero amplitude.

**Figure 4**    (Color online) Cationic phonon displacements for three low-energy phonon modes: $\omega_2$ (a), $\omega_3$ (b), $\omega_4$ (c). The phonons were calculated for the *B1a1* lattice under 0.04% transverse strain. $\omega_1$ is not pictured as it is characterized by transverse displacement of fluorite-like and perovskite layers along the non-polar $a$ axis and negligible displacement along the polar axes.

**Figure 5**    (Color online) The dipolar contributions per layer of fluorite-like and perovskite-like layers. The fluorite-like layer contribution is larger for the hard modes $\omega_1$ and $\omega_4$, while the perovskite-like layer contribution is larger for the soft modes $\omega_2$ and $\omega_3$ (see text for detailed discussions).

**Figure 6**    (Color online) Atom resolved density of states for P- (a), As- (b), and Sb-doped (c) BiT and total density of states (TDOS) for the doped system compared to pure BiT (d)-(f). The dopant atoms show stronger hybridized. Stated from the dopant atom in the conduction band reduced the effective band gap for P and As.

**Table I.** Components of the BEC tensor for fully relaxed, equilibrium *B1a1* BiT for as a function of ion and sublattice [perovskite (P)/fluorite (F)]. The equivalent atoms are primed.

| Atom type | Sublattice | Born effective charge | | |
|:---:|:---:|:---:|:---:|:---:|
| | | $Z^*_x$ | $Z^*_y$ | $Z^*_z$ |
| Bi1 | PL | 5.27 | 4.57 | 4.44 |
| Bi1′ | PL | 5.35 | 5.04 | 4.28 |
| Bi2 | FL | 4.74 | 4.74 | 4.87 |
| Bi2′ | FL | 4.76 | 4.77 | 4.65 |
| Ti1 | PL | 7.02 | 5.74 | 5.42 |
| Ti2 | PL | 6.47 | 5.53 | 5.70 |
| Ti2′ | PL | 6.20 | 5.49 | 5.58 |
| O1 | PL | −4.12 | −3.30 | −2.07 |
| O1′ | PL | −3.72 | −3.64 | −2.01 |
| O2 | PL | −2.89 | −2.78 | −3.66 |
| O2′ | PL | −4.29 | −2.89 | −2.34 |
| O3 | PL | −3.99 | −3.70 | −2.03 |
| O3′ | PL | −2.89 | −2.84 | −3.65 |
| O4 | PL | −3.73 | −3.68 | −2.04 |
| O4′ | PL | −4.15 | −3.22 | −2.07 |
| O5 | FL | −2.17 | −2.05 | −4.71 |
| O5′ | FL | −2.96 | −2.98 | −2.76 |
| O6 | FL | −2.97 | −2.98 | −2.80 |
| O6′ | FL | −1.96 | −1.85 | −4.80 |

**Table II.** The average off-center displacements of charge centers (ODC) between the cation atom and the nearest neighbor oxygen atoms and the Born effective charges (BEC) for selected site-specific cations.

| $B1a1$ | No. O neighbor | ODC$_a$ | ODC$_b$ | ODC$_c$ | BEC$_a$ | BEC$_b$ | BEC$_c$ | Nominal charge |
|---|---|---|---|---|---|---|---|---|
| Bi$_{\text{Fluorite-like}}$ | 8 | 0.0 | 0.387 | −0.263 | 5.310 | 4.803 | 4.363 | +3 |
| Bi$_{\text{Perovskite-like}}$ | 6 | 0.0 | 0.202 | 0.032 | 4.750 | 4.758 | 4.762 | +3 |
| Ti$_{\text{Perovskite-like}}$ | 6 | 0.0 | −0.179 | 0.032 | 6.566 | 5.585 | 5.566 | +4 |

**Table III.** Components and absolute value of site and layer resolved dipole moment LRDM (%) (see Eq. (3)) for the four lowest optical phonon modes. The energy eigenvalues are in the order $\omega_1 < \omega_2 < \omega_3 < \omega_4$.

| Phonon frequency | Site type | $(LRDM)_a$ | $(LRDM)_b$ | $(LRDM)_c$ | abs(LRDM) |
|---|---|---|---|---|---|
| $\omega_1$ | Bi | 1.133 | −0.060 | 0.807 | 1.395 |
| | Ti | 0.012 | −0.495 | 0.140 | 0.515 |
| | O | −0.145 | 1.555 | 0.052 | 0.218 |
| | Fluorite layer | 1.094 | 1.378 | 0.833 | 1.947 |
| | Perovskite layer | −0.108 | 0.116 | 0.027 | 0.161 |
| $\omega_2$ | Bi | −0.426 | −2.635 | −1.042 | 2.865 |
| | Ti | −1.376 | −6.071 | −0.722 | 6.267 |
| | O | 2.802 | 9.706 | 0.230 | 10.105 |
| | Fluorite layer | −0.255 | 0.869 | −0.596 | 1.084 |
| | Perovskite layer | 2.631 | 6.176 | 2.326 | 7.105 |
| $\omega_3$ | Bi | −0.293 | 1.755 | −0.133 | 1.784 |
| | Ti | −0.961 | −4.151 | −0.381 | 4.278 |
| | O | 2.254 | 3.396 | 1.514 | 4.348 |
| | Fluorite layer | 0.148 | −0.355 | −0.238 | 0.452 |
| | Perovskite layer | 2.107 | 5.519 | 1.619 | 6.125 |
| $\omega_4$ | Bi | −0.343 | −1.922 | 3.583 | 27.316 |
| | Ti | −1.737 | 0.448 | −11.274 | 4.080 |
| | O | 0.959 | 2.475 | 8.690 | 11.416 |
| | Fluorite layer | 0.320 | 2.202 | −15.0 | 15.164 |
| | Perovskite layer | 2.417 | −1.641 | 27.316 | 27.472 |

**Table VI.** Computed values of components of spontaneous polarization and its norm (µC/cm$^2$), polar angle (degrees) and band gap (eV) for epitaxially strained and doped BiT.

|            | $P_a$  | $P_b$   | $P_c$   | $P$     | Polar angle | Band gap |
|------------|--------|---------|---------|---------|-------------|----------|
| Unstrained | 0.0    | 52.255  | 7.879   | 52.847  | 81.425      | 2.173    |
| P doped    | 2.642  | 47.955  | 35.031  | 59.445  | 53.852      | 1.386    |
| As doped   | 2.923  | 45.806  | 23.523  | 51.577  | 62.817      | 1.793    |
| Sb doped   | 5.795  | 39.738  | 21.014  | 45.325  | 62.129      | 2.217    |

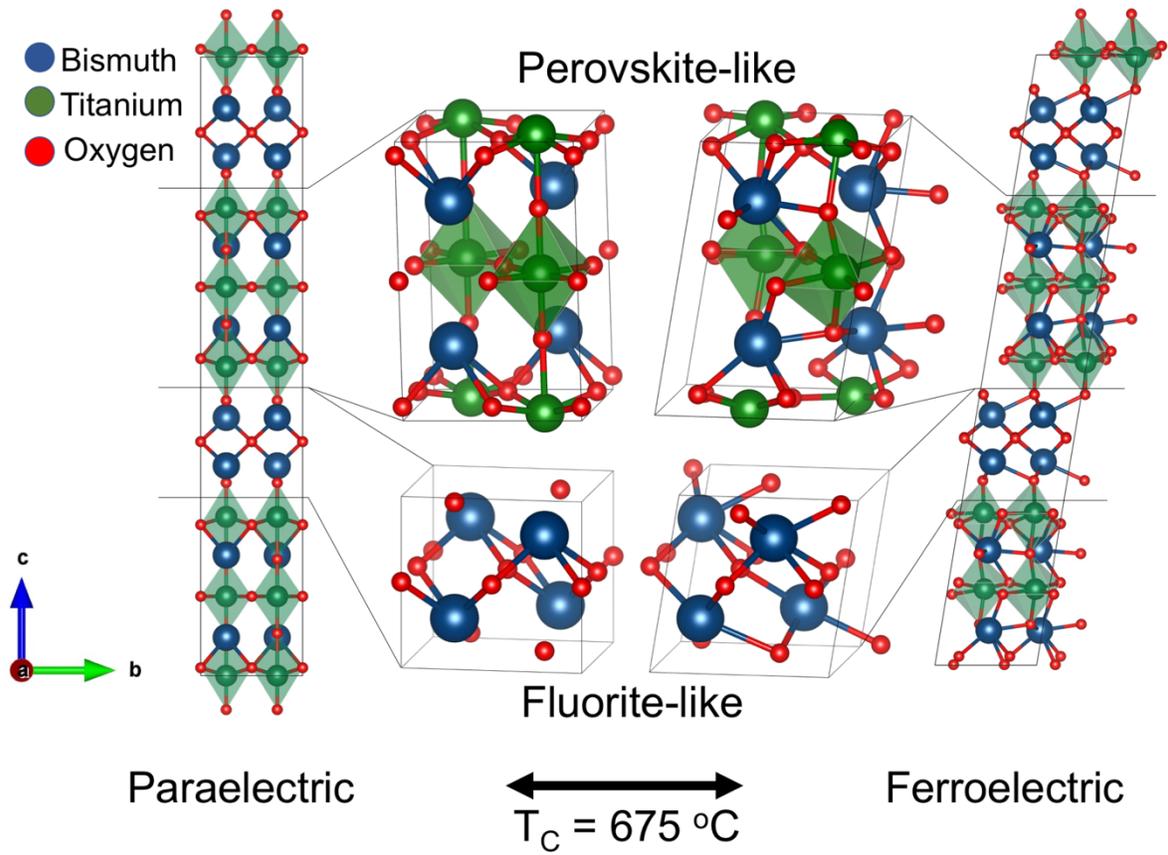

Figure 1

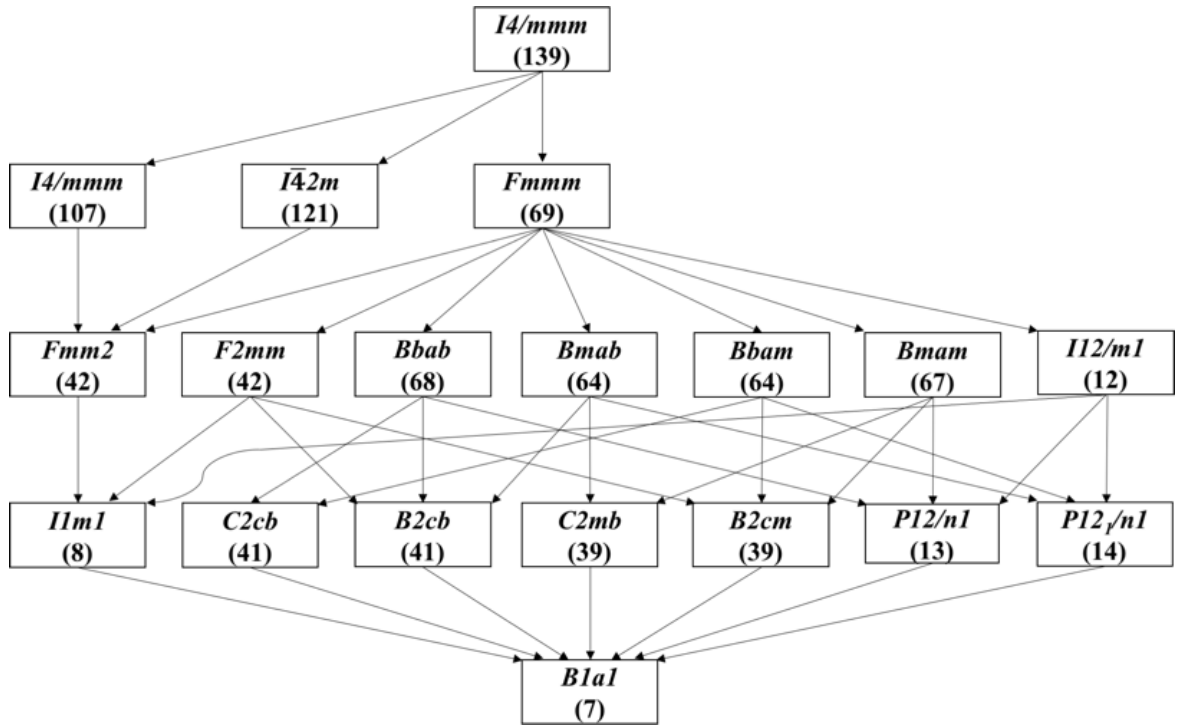

**Figure 2**

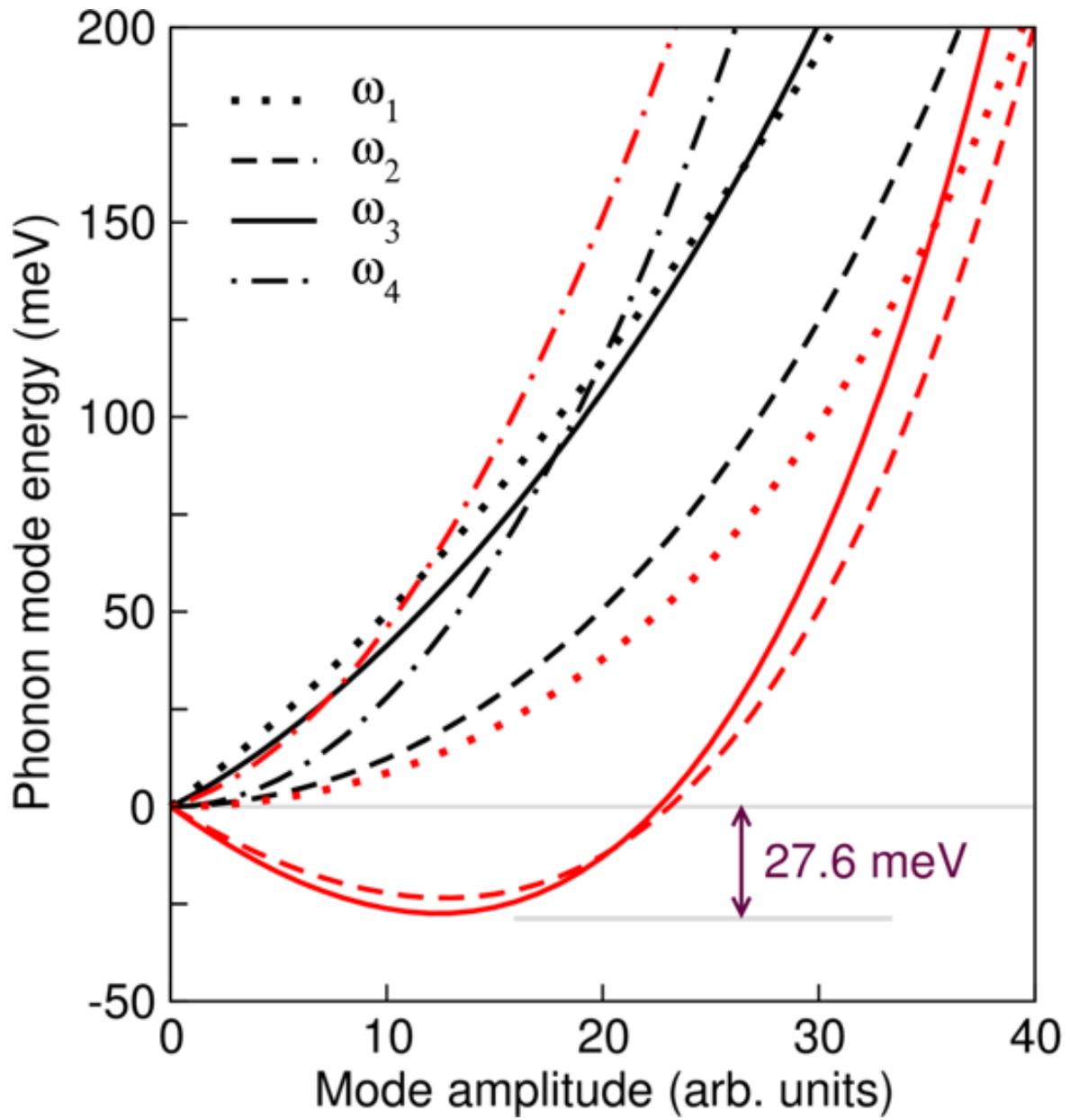

**Figure 3**

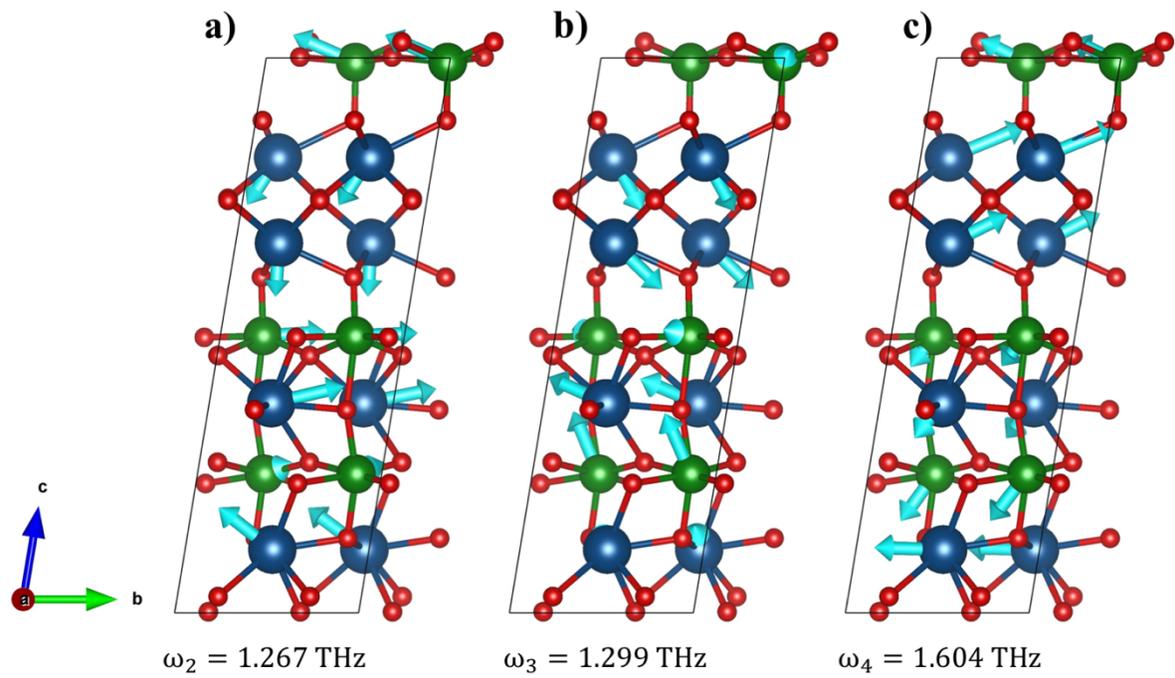

$\omega_2 = 1.267$ THz $\qquad$ $\omega_3 = 1.299$ THz $\qquad$ $\omega_4 = 1.604$ THz

**Figure 4**

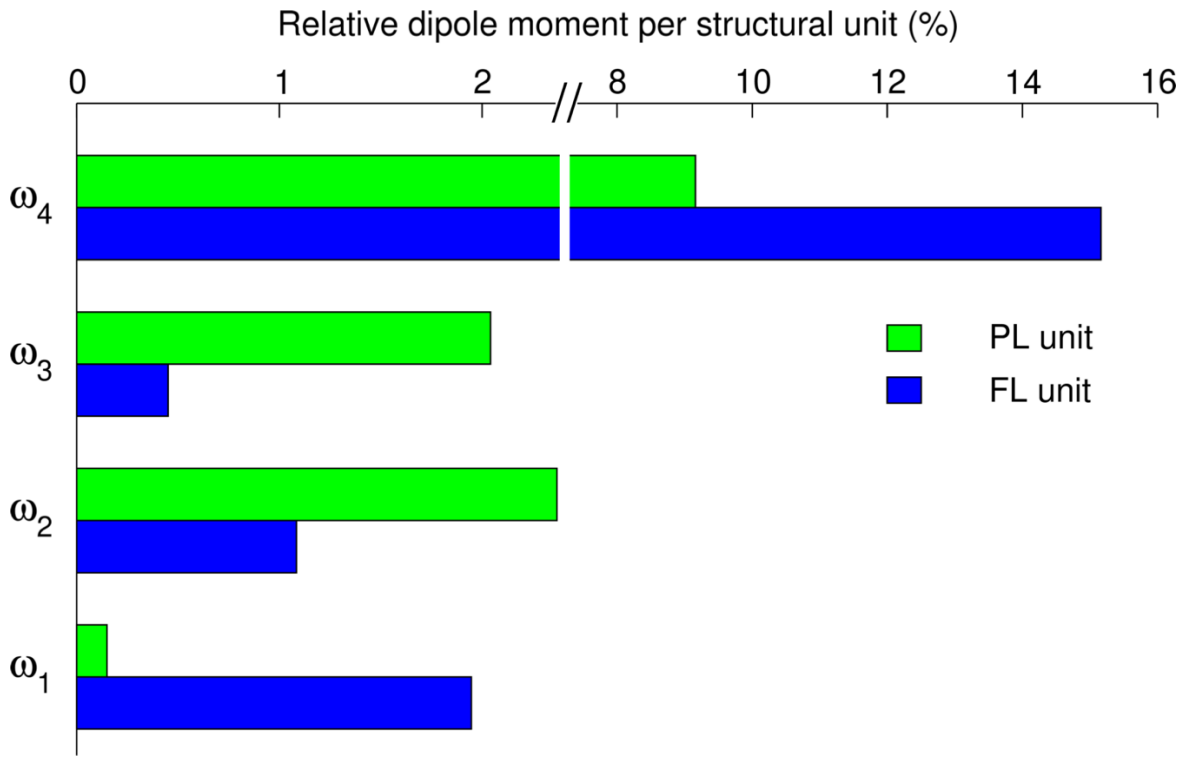

**Figure 5**

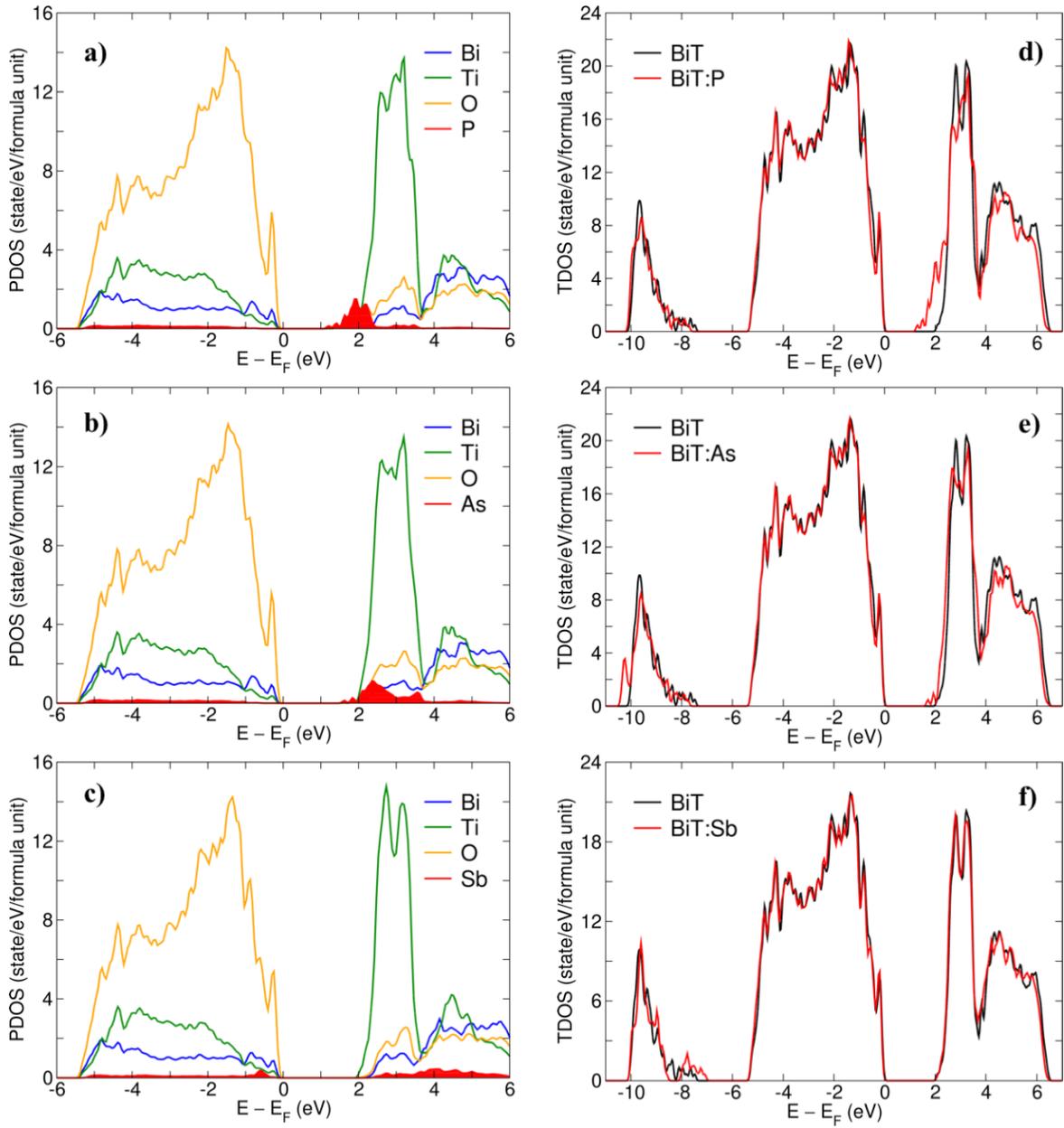

**Figure 6**